\documentclass[prl,aps,footinbib,twocolumn]{revtex4}
\usepackage{amssymb}
\usepackage{amssymb}
\usepackage{graphicx}
\usepackage{amsmath}
\usepackage{times}
\usepackage{color}
\usepackage{subfigure}
\usepackage{setspace}
\usepackage{bm}

\begin{document}

\title{ Resolved atomic interaction sidebands in an optical clock transition }

\author
{ M. Bishof$^{1}$, Y. Lin$^{1}$, M. D. Swallows$^{1}$,  A. V. Gorshkov$^{2}$, %\\
 J. Ye$^{1}$ and A. M. Rey$^{1}$ }
\affiliation{$^{1}${ JILA and Department of Physics, NIST and University of Colorado, Boulder, CO 80309-0440, USA}}
 \affiliation{$^{2}$Institute for Quantum Information, California Institute of Technology, Pasadena, CA 91125, USA}

 %\affiliation{$^\ast$To whom correspondence should be addressed; E-mail: arey@jilau1.colorado.edu}

\date{\today}
\begin{abstract}
We report the observation of resolved  atomic interaction sidebands (ISB) in the ${}^{87}$Sr optical clock transition when atoms at microkelvin temperatures
are confined in a two-dimensional (2D) optical lattice. The ISB are a manifestation of the strong interactions that occur between atoms confined in a quasi-one-dimensional geometry and disappear when the confinement is relaxed along one dimension.  The emergence  of ISB is linked to  the recently observed suppression of  collisional frequency shifts in \cite{Swallows11}. %M. D. Swallows \textit{et al.} \textit{Science}, 3 February 2011.
 At the current   temperatures, the ISB can be resolved  but are broad. At lower temperatures, ISB are predicted  to  be substantially narrower and usable as  powerful  spectroscopic tools in strongly interacting alkaline-earth gases.
\end{abstract}
 \maketitle

{ Experimental efforts in control of ultracold alkaline-earth
 atoms (AEA), such as Sr or Yb, have led to remarkable developments in optical atomic clocks
 ~\cite{Swallows11,NISTYb}, which are approaching the accuracy of single ion standards \cite{Rosenband}}. Fermionic AEA  are also beginning to attract considerable  attention in the context of quantum information processing \cite{Gorshkov2009} and quantum simulation
\cite{expe2009,Gorshkov20092}. Many  of these  applications   require  reaching the strongly interacting regime,  in which interactions dominate over all relevant dynamical energy scales.
In this Letter, we demonstrate the onset of strong interactions
in a low dimensional  gas of ${}^{87}$Sr atoms by observing resolved interaction sidebands (ISB) in Rabi spectroscopy of the  ${}^1S_0-{}^3P_0 $  %(${}^1S_0$ to ${}^3P_0 $)
optical clock transition.
%In our experiment, interactions are enhanced by tightly confining the atoms in a two %dimensional optical lattice and we observe that the ISB disappear when the atoms are %confined in a one dimensional lattice.

 %At  $T\sim 4-5 \mu$K.  At this temperature,  in the 2D lattice gas with sites  populated by mainly one  and two atoms,

{ Rabi spectroscopy has served as an excellent probe for interacting alkali gases. For example, it was used to detect and characterize Bose Einstein condensation in spin polarized hydrogen~\cite{Killian}, to directly measure the fermion pair wave function in the BEC-BCS crossover~\cite{Rf}, and to resolve the  Mott insulator shell structure in a bosonic lattice system~\cite{Campbell06}. Here, we show that the ISB structure that emerges during optical Rabi interrogation can become a powerful spectroscopy tool to uniquely probe lattice-trapped AEA. At microkelvin temperatures and loading conditions of mainly one or two atoms  per  lattice site,  we observe a single  ISB, which  can be resolved from the carrier but is broad. However, in the quantum degenerate regime, the ISB are expected to become substantially  narrower and could thus lead to a precise determination of the  $s$-wave ${}^1S_0-{}^3P_0 $ interaction parameters (currently, the ${}^1S_0-{}^3P_0$ scattering lengths for fermionic
AEA remain unknown). In addition, we show that the variation in ISB spectra
between sites with different numbers of atoms allows for a precise
characterization of atom number occupations in interacting AEA
systems.}

 \begin{figure}
{\includegraphics[width =0.7\linewidth]{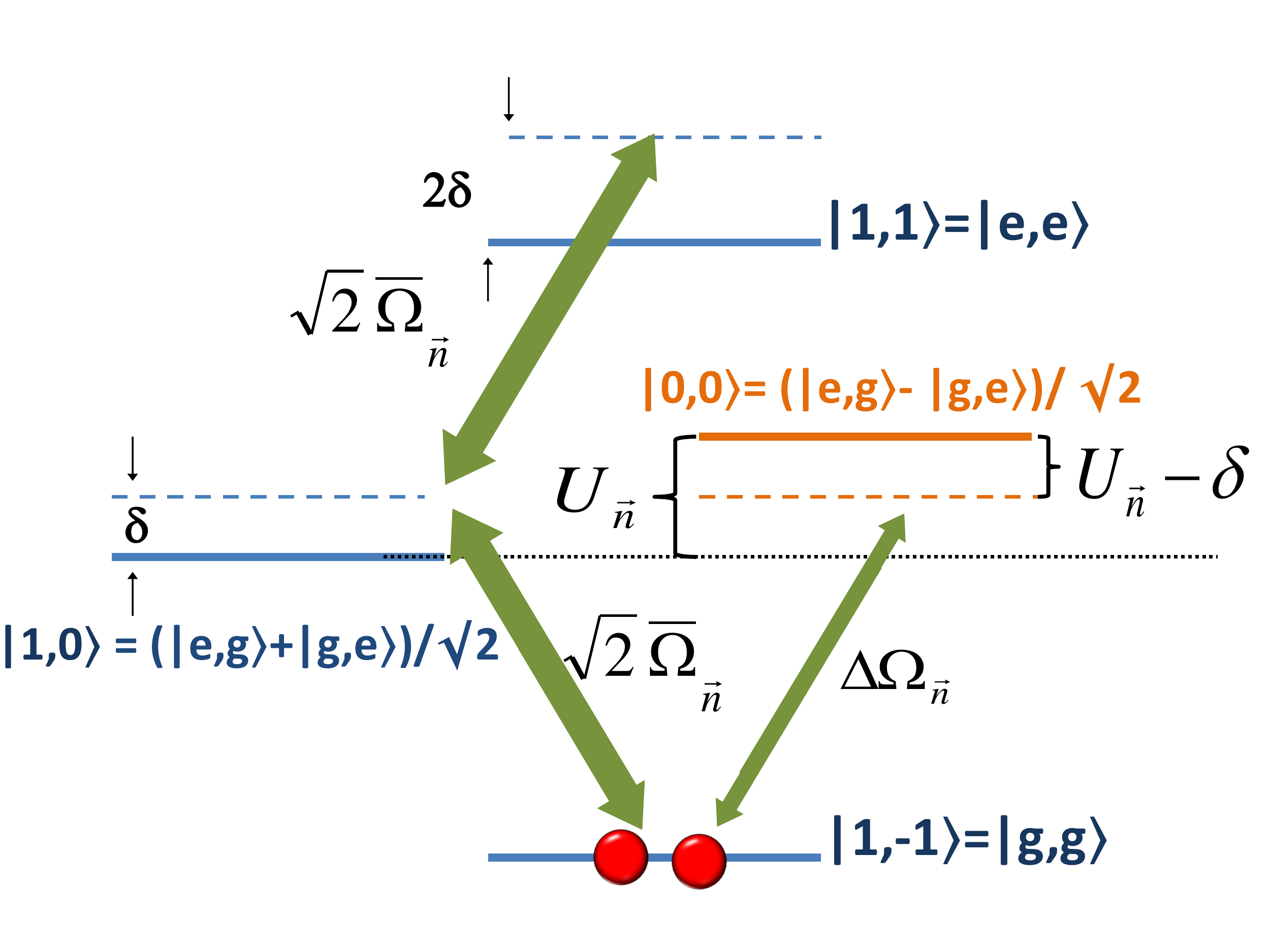}}
 \leavevmode
 \caption{ {(Color online) Schematic diagram of Rabi
spectroscopy in the strongly interacting regime. When $\delta \approx
0$, only the non-interacting triplet states ($S=1$) are accessible,
giving rise to an interaction-free carrier. However, when $\delta \sim
U_{\vec{n}}$, the singlet $|0,0\rangle$ becomes resonant, and a
slightly inhomogeneous coupling ($\Delta \Omega_{\vec{n}} \neq 0$) can
induce population transfer to it, as manifested in an ISB.}}  \label{fig1}
\end{figure}

When the ${}^1S_0$ and ${}^3P_0$ electronic degrees of freedom within a single fermionic atom are represented as an  effective pseudo-spin 1/2  ($g$ and $e$), two nuclear-spin polarized atoms form a simple model for understanding the appearance of ISB (see Fig.\ \ref{fig1}). We consider atoms trapped in a tube geometry  with a weak trapping frequency  $\omega_Z$ along  $\hat Z$  and a strong  transverse confinement $\hbar \omega_{X,Y}\gg k_B T_{X,Y}$, large enough to forbid any transverse dynamics [$T_X(T_Y)$ is the temperature along $\hat X (\hat Y)$].  We assume the atoms are initially prepared in the same pseudo-spin   state ($|gg\rangle$) and interrogated   by  a linearly polarized laser beam with bare Rabi frequency $\Omega^B$ and detuning from the atomic resonance $\delta$.  %=\omega_L-\omega_0$.
The Pauli exclusion principle forces   atoms in identical internal  states  to occupy  different vibrational levels along $\hat{Z}$, $\vec{n}=(n_1,n_2)$, since their spatial wave function must be antisymmetric. %Thus
Being identical fermions, the atoms initially
experience zero $s$-wave interaction. If  the atoms are
coherently driven to $e$ (i.e.\ if they experience the same Rabi frequency $\bar{\Omega}_{\vec{n}}\equiv (\Omega_{n_1}+\Omega_{n_2})/2$), then the effective spin must remain symmetric under exchange. Consequently, during the excitation the atoms will not experience any $s$-wave interactions.  We refer to states with spin-symmetric wave functions, $|gg\rangle \equiv |S = 1, M = -1\rangle$, $|ee\rangle \equiv |1, 1\rangle$ and $(|eg\rangle+|ge\rangle)/\sqrt{2} \equiv |1,0\rangle$,  as the triplet states, $S=1$ {\cite{Rey09,Gibble09}}.
%Since they do not interact their dressed energy levels correspond to $\pm\sqrt{\bar{\Omega}_{\vec{n}}^2+\delta^2}$ and $0$ .

However, in optical transitions, a small component of the probe beam along $\hat{Z}$ leads to a slightly different Rabi frequency $\Omega_{n_j}(\eta_Z^2)$ for each mode, with $\eta_{Z}=k_{Z}a^{Z}_{ho}/ \sqrt{2}$ the Lamb-Dicke parameter, $m$ the atom mass,
$k_{Z}$  the component of the probe laser wave vector along $\hat{Z}$ and   $a^{Z}_{ho}= \sqrt{\frac{\hbar} {m\omega_{Z}}}$  the  $\hat{Z}$ harmonic oscillator length. If $\Delta \Omega_{\vec{n}}=(\Omega_{n_1}-\Omega_{n_2})/\sqrt{2}$ is not zero, then the optical excitation-induced inhomogeneity can transfer some of the atoms to the antisymmetric pseudo-spin singlet state, $(|eg\rangle-|ge\rangle)/\sqrt{2} \equiv |0,0\rangle$.  The interacting singlet %state
is separated from %the triplet states
$|1,0\rangle$ %(|eg\rangle+|ge\rangle)/\sqrt{2}$
by an interaction energy $U_{\vec{n}}$.

In the limit where $\delta\sim \bar{\Omega}_{\vec{n}}\ll U_{\vec{n}}$, a  small  $\Delta \Omega_{\vec{n}}$ cannot  overcome the energy cost required to  drive the transition to the singlet (see Fig.\ \ref{fig1}). Evolution into the singlet is blocked and  the atoms remain non-interacting.  However, if we increase the detuning to $\delta \sim U_{\vec{n}}$, the transition $|1,-1\rangle \rightarrow |0,0\rangle$ %$|gg\rangle \rightarrow (|eg\rangle-|ge\rangle)/\sqrt{2}$
becomes resonant.
%the  initially occupied dressed state (which is almost a pure  $|gg\rangle$) can be made  resonant with the singlet, which is separated from $|gg\rangle$ by an  energy  $\sim U_{\vec{n}}-\delta$).
Here,  a small $\Delta \Omega_{\vec{n}}$ can  efficiently %again
transfer the atoms to the singlet, giving rise  to an ISB in the lineshape. Note that if  instead of preparing all the atoms  in  $g$ we prepare the atoms in $e$, the ISB  occurs at $\delta \sim -U_{\vec{n}}$. %since the singlet is separated from $|ee\rangle$ by an energy $\delta + \bar{U}_{\vec{n}}$.

The ISB resemble motional sidebands that develop in spectroscopy of trapped ions %atoms
when the trapping frequency is much larger than the recoil energy, the so called Lamb-Dicke regime \cite{Wineland1979}. In this case, the carrier becomes free of Doppler shifts up to the corresponding ``Stark shift'' \cite{Blatt2003}, and all motional effects are manifested  in the sidebands.  In our case, the  ISB are pushed away from the carrier due to the  large energy separation between interacting and non-interacting states, leaving the carrier  free from interaction effects, up to a correction proportional to $\Delta \Omega_{\vec{n}}^2/U_{\vec{n}}$. Recently we have observed the suppression of collisional frequency shifts to the $10^{-17}$ level under similar tight trapping conditions \cite{Swallows11}.  The observation of ISB unambiguously confirms  this suppression to be the  consequence of reaching the strongly interacting regime.

Our experiment employs nuclear spin-polarized ($m_I= +9/2$) ${}^{87}$Sr atoms loaded in  two
lattice configurations: a one-dimensional (1D) vertical lattice which  creates  an array of quasi-two-dimensional traps (pancakes) along  $\hat Y$ with  weak confinement along  the $\hat X$ and $\hat Z$ directions, and  a deep 2D lattice which creates an array of quasi-1D traps (tubes) in the $\hat X$-$\hat Y$ plane and has weak confinement along $\hat Z$.
To prepare the atomic system, we laser cool ${}^{87}$Sr atoms to about 2 $\mu$K inside a magneto-optic trap based on the weak ${}^1S_0-{}^3P_1$ transition and then load them into a 1D vertical lattice. The spatial distribution of occupied 1D lattice sites is determined by the vertical extent of the MOT cloud, which is approximately Gaussian with a standard deviation $\sigma_V = 30$ $\mu$m. For the 2D lattice confinement  we then adiabatically ramp up the horizontal lattice (along $\hat X$).  To remove any atoms
trapped in the 1D vertical lattice outside the 2D intersection region, we ramp the vertical
lattice off and then back on. The number of horizontal lattice sites occupied is
determined by the radial temperature of the vertical lattice. Atoms load into $\sim$100 rows of tubes uniformly distributed along $\hat Y$, while the "columns" distributed along $\hat X$ are
loaded according to a Gaussian distribution with standard deviation $\sigma_H$ of 6-10 $\mu$m.
After forming the lattices, we perform Doppler and sideband cooling using the ${}^1S_0-{}^3P_1 $, $ F= 11/2$ transition. Simultaneously, atoms are optically pumped to the $m_I = +9/2$
ground state magnetic sublevel, using $\sigma^+$-polarized light on the %${}^1S_0-{}^3P_1$; $F = 9/2$
${}^1S_0-{}^3P_1$, $F = 9/2$ transition,
propagating %directed
along a bias magnetic field parallel to  $\hat Z$. We perform spectroscopy of the ${}^1S_0-{}^3P_0$ transition using
a narrow linewidth laser propagating along $\hat Y$.
%The  ${}^1S_0-{}^3P_1$ spectroscopy  is performed by  using  a 80-ms pulse. The laser intensity is scanned to achieve maximum excitation fractions and %from it we estimate $\langle {\bar\Omega} \rangle_{T_Z}\sim  2\pi \times 6.25$ Hz.

We quantify the number of atoms
by detecting fluorescence on the strong ${}^1S_0-{}^1P_1$ transition at 461 nm. Temperature information and trap frequencies are extracted  from Doppler spectroscopy along $\hat Z$ and vibrational sideband spectra \cite{Blatt09,Swallows11}. We estimate that we loaded  $\sim$7000 atoms in the 1D lattice distributed in $\sim$420 pancakes and $\sim$17 atoms per pancake.  We determine $\omega_Y=2\pi \times 80 $ kHz and $\omega_{X,Z}=2\pi \times 500 $ Hz, and    $T_{X,Y,Z}\sim 4$ $\mu$K. For the 2D lattice, we loaded $\sim$3000 atoms and estimate  that 30$\%$ of the populated tubes are doubly occupied. We determine   $\omega_X = 2 \pi \times 110$ kHz, $\omega_Y=2\pi \times 70$ kHz and $\omega_Z=2\pi \times 800$ Hz at trap center, and $T_{X,Y,Z}\sim 4.5$ $\mu$K.
%Our current transverse temperature is a factor of two larger than the one  in  prior  experiment in which the clock shift suppression was observed %\cite{Swallows10}.  Nevertheless even at the current experimental conditions we expect  tunneling  between tubes  to be  suppressed by the  energy %offset between adjacent lattice sites generated by  the Gaussian beam profile.

ISB develop in the lineshape only in the parameter regime  where $\gamma\equiv E_{\rm int}/\langle {\bar\Omega} \rangle_{T}\gg 1$. Here, $E_{\rm int}=(N-1)|\langle U \rangle_{T}|/2$ is the mean interaction energy per particle and  $N$  represents the number of atoms per lattice site.
 $\langle U \rangle_{T}$ characterize the strength of two-body interactions and depends both on the confinement volume and temperature as \cite{supp}:   $\langle U \rangle_{T}\approx u \vartheta(\frac{\hbar\omega_X }{ k_B T_X})\vartheta(\frac{\hbar\omega_Y }{ k_B T_Y})\tilde\vartheta(\frac{\hbar\omega_Z }{ k_B T_Z})$ with   $u = 4  a_{eg}^- \sqrt{\frac{m \omega_X \omega_Y \omega_Z}{h}}$, $a_{eg}^-$  the singlet ${}^1S_0-{}^3P_0$ scattering length \cite{Gorshkov20092}  and $\vartheta$ and  $\tilde\vartheta$  functions characterizing  the temperature dependence, i.e.  $\vartheta(x \gg 1)\sim 2\tilde\vartheta(x\gg 1)\to 1$  and $\vartheta(x\ll 1)\sim \tilde\vartheta(x\ll 1)\to \sqrt{x}$. We used $\langle O \rangle_{T} $ to denote the thermal average of  $O$. In our experiment, the  ${}^1S_0-{}^3P_0$ spectroscopy  is generally  performed by  using  a  80-ms pulse and the ``$\pi$-pulse'' laser intensity is set to achieve maximum excitation fraction. We estimate $\langle {\bar\Omega} \rangle_{T}\sim  2\pi \times 6.25$ Hz.

\begin{figure}[htbp]
\includegraphics[width =1\linewidth, trim=0.15in 0in 0.15in 0.1in]{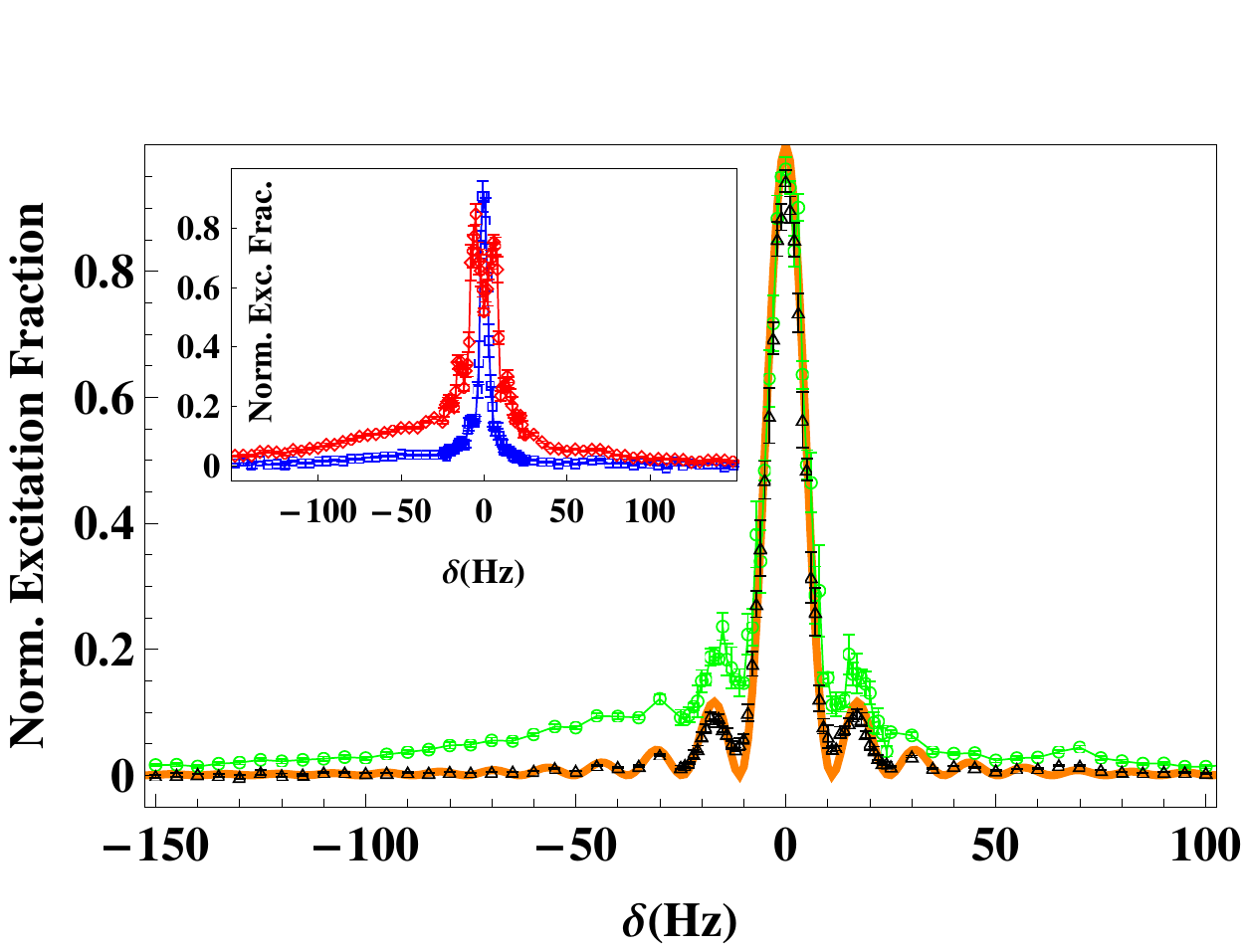}
 \leavevmode
 \caption{{ (Color online) Measured Rabi lineshapes for $g$ to $e$ interrogation (normalized with respect to peak height).  The main Panel shows lineshapes taken in a 1D lattice (black triangles) and in a 2D lattice (connected green circles)
  when the transition is excited with a $\pi$-pulse with 80 ms duration. The noninteracting, homogeneous Rabi lineshape for an 80 ms $\pi$-pulse is also shown (orange line).  The inset shows lineshapes measured in a 2D lattice using a 160 ms $2\pi$-pulse (connected red diamonds) and a 160 ms  $\pi$-pulse (connected blue squares).  Each lineshape was measured under the trapping and temperature conditions described in the main text.  The data from each lineshape is collected into bins 1-5 Hz wide.  }}
   \label{expe}
\end{figure}

Using the experimental trapping conditions and temperatures and assuming a moderate $a_{eg}^-=-70 a_0$ ($a_0$ the Bohr radius), we obtain  $\gamma^{2D}\sim 10$  in the doubly occupied sites at the center of  the 2D lattice. %when we choose a moderate   $a_{eg}^-=-70 a_0$.
On the contrary, for the central pancake of the 1D lattice with a mean number of 17 atoms and the same scattering length,  $\gamma$ is reduced to
$\gamma^{1D}\sim 0.6$. Based on these values, one predicts the development of ISB  only in the 2D lattice geometry, a prediction  confirmed by our measurements.  We measured  a series of  lineshapes in the 1D and 2D lattices. In the 2D lattice case  we  also varied the pulse area and Rabi frequency.  The resulting data are shown in Fig.\ \ref{expe}.
To remove some of the statistical noise we performed multiple scans across the transition resonance under each experimental condition.  For each scan, the drift of our ultra stable laser was canceled to $<$50 mHz per second, and the laser was stepped by 2 Hz each experimental cycle ($\sim$1.5 sec).  The center of each scan was determined by a fit to a Lorentzian.  Each data set consists of $\sim$20 concatenated scans and the direction of the scan was alternated to reduce systematic effects due to residual drift.  We separated the data into bins and calculated the mean and standard error of the mean for each bin.  At our $4-5$ $\mu$K temperatures, the ISB  in the 2D  lattice are noticeable but nevertheless broad. To improve  resolution, we reflect the binned lineshape at positive  detuning (free of ISB %interaction sideband
when  $a_{eg}^-<0 $)  and subtract it from the corresponding bin at negative detuning.  The resultant lineshapes with a cleaner ISB  are shown  in Fig.\ \ref{lineshape}.

The presence  of ISB in the tightly confined geometry can be understood quantitatively by considering an isolated tube with $N$ atoms. We characterize the tube dynamics by defining a set of effective spin operators, $S^{x,y,z}_{n_j}$, in the $\left\{e,g\right\}$ basis~\cite{Swallows11,Rey09}. Here the subscript $n_j$ is drawn from %specifies
a fixed set  $\vec{n}=\{n_1,n_2 \dots n_N\}$ of initially populated vibrational modes along $\hat Z$.  The description of the system in terms of effective spin operators is valid provided those initially populated modes remain singly occupied by either a $g$ or an $e$  atom during the excitation process.  In the rotating frame, the Hamiltonian of the system  up to constant terms becomes~\cite{Rey09},
\begin{eqnarray}
\frac{\hat H^{S}_{\vec{ n}}}{\hbar} = -\! \delta {S}^z \!- \!\sum_{j=1}^N{\Omega_{n_j} {S}^{x}_{n_j}}\!
- \!  \!\sum_{ j'\neq j }^N{\frac{ U_{n_j ,n_{j'}} }{2}(\vec{ S}_{n_j}\cdot \vec{ S}_{ n_{j'}}-\frac{1}{4}) }. \label{many}
\end{eqnarray} $S^{z,x} = \sum_{j=1}^N {S}^{z,x}_{n_j}$ are collective spin operators. The quantity
%$U_{n_j ,n_{j'}}= 4  a_{eg}^- \sqrt{m \omega_X \omega_Y \omega_Z/\hbar} \vartheta(\frac{\hbar \omega_X}{k_B T_X})  \vartheta(\frac{\hbar \omega_Y}{k_B T_Y})I_{n_{j},n_{j'}}$
$U_{n_j ,n_{j'}}= u \vartheta(\frac{\hbar \omega_X}{k_B T_X})  \vartheta(\frac{\hbar \omega_Y}{k_B T_Y})  I_{n_{j},n_{j'}}$
measures the strength of the interactions between two atoms in the antisymmetric electronic state.  $I_{{n_j},{n_{j'}}}$ is an overlap integral between harmonic oscillator modes along $\hat Z$ \cite{supp}. Only the $\hat Z-$mode distribution is treated exactly since   at current temperatures only a few transverse excited modes are populated. The population of transverse  modes is  accounted for  as a  renormalization of the interaction parameter. %is a mode overlap coefficient characterizing the temperature dependence along $T_Z$ \cite{supp}.}
 \begin{figure}[htcp]
{\includegraphics[width =1.\linewidth]{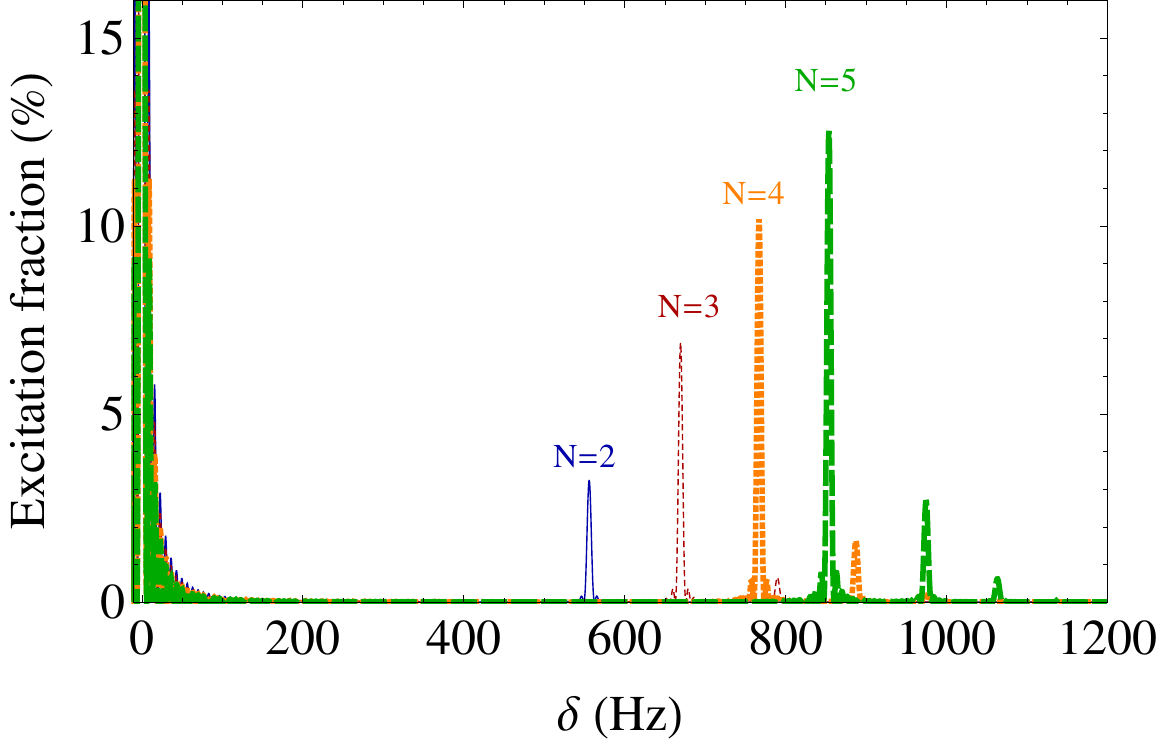}}
 \leavevmode
 \caption{ {(Color online) Excitation fraction vs. $\delta$ at $T=0$  for lattice sites with $N=2-5$ atoms. The ISB reveal the $S=N/2-1$ spectrum and can be used to probe occupation number in a lattice. The parameters used for the plot were $\eta_Z=0.4$,  $\Omega^B= 2 \pi \times 5$ Hz,  $t=1.5 \pi/\Omega^B$ and $u=2 \pi \times 2800$ Hz. }
 } \label{fig3}
\end{figure}

The  interaction part of the Hamiltonian is diagonal in the collective spin basis $|S,M\rangle$, $S=0(\frac{1}{2}),\dots, N/2$ and $|M|\leq S$. For $N=2$, the spin basis is spanned by the triplet states $\left|1,M\right\rangle$ and the singlet $\left|0,0\right\rangle$. Among the collective states only the $S=N/2$ states are noninteracting. States with $S<N/2$ experience a finite interaction energy.  In the presence of excitation inhomogeneity, $S$ is no longer conserved and during excitation of the clock transition, atoms can  be transferred mainly between $S=N/2$ and $S=N/2-1$ states. { The $N-1$ collective
excitation modes $|S=N/2-1,M=N/2-1,q\rangle$ ($q=1,\dots N-1$) with
energies ${U}_{\vec{n}}^{q,N}$ can be accessed from the  initially
$g$-polarized state  $|N/2,-N/2\rangle$  using $\delta \sim
{U}_{\vec{n}}^{q,N}$. They give rise to $N-1$ sidebands in the
lineshape (Fig.\ \ref{fig3}): \begin{eqnarray}
 N^{e}_{\vec{n}}(t,\delta) &\approx&  N f(t,\delta, \bar{\Omega}_{\vec{n}})+ \sum_{q=1}^{N-1}f(t,\delta -{U}_{\vec{n}}^{q,N}, \Delta{\Omega}_{\vec{n}}^{q,N}), \label{R}
\end{eqnarray} with $f(t,\delta, y)\equiv  \frac{y^2}{y^2+\delta^2} \sin^2\left (\frac{t \sqrt{ y^2+\delta^2}}{2}\right)$ and  $\Delta{\Omega}^{q,N}_{\vec{n}}=2 \sum_j \Omega_{n_j} \langle N/2,N/2|S^x_j |N/2-1,N/2-1,q\rangle$. The first term is associated with the carrier and the second with the ISB. For the case $N=2$, we have ${U}_{\vec{n}}^{1,2}={U}_{\vec{n}}$ and  $\Delta{\Omega}^{1,2}_{\vec{n}}=\Delta{\Omega}_{\vec{n}}$.

At quantum degeneracy, only the lowest lying vibrational modes need to be considered $\vec{n}=\{0,1,\dots N-1\}$.  In this case, the interaction sidebands are very narrow. Assuming  $t=s \pi/\Omega^B$, with $s$ a numerical constant that describes the pulse duration, the peak width is just $2\Omega^B/s$. Measuring the interaction sidebands thus precisely determines $a_{eg}^-$  %the corresponding singlet  $S-P$ scattering length
and characterizes the $S=N/2-1$ spectrum, which has structure for $N>2$.  Since  the  peak height of the interaction sidebands scales as  $\eta_Z^4$,  large probe inhomogeneities  are required to observe them at low temperature.

In Ref.~\cite{Campbell06}, interaction-dependent transition frequency
shifts in  RF spectroscopy were used to spectroscopically distinguish sites with different occupation numbers, revealing the shell structure of the Mott insulator phase.  The $g$ and $e$ states in  AEA have no hyperfine structure, and hence RF spectroscopy is not applicable. However, as shown in Fig.\ \ref{fig3}, the optical analog of RF spectroscopy-- the ISB
spectroscopy -- can distinguish sites with different atom numbers. This capability can have important implications in quantum simulation~\cite{Gorshkov20092} and information proposals~\cite{Gorshkov2009} with AEA. }

 At  finite $T$, instead of a fixed set of populated modes, a thermal average should be taken. For simplicity  and to correspond to our current experiment, we focus on the $N=2$ case. After evaluating the thermal average \cite{supp}, we obtain the following
approximate expression for the ISB, valid in the regime  $|\delta| \gg \Omega^B$,  $k_B T_Z>\hbar \omega_Z$ and  $t=s\pi/\Omega^B   \ll \langle \Delta \Omega\rangle_{T_Z}^{-1}$ (These conditions are satisfied  in our experiment):
\begin{eqnarray}
\langle \hat{N}^e \rangle_{T} \approx    \frac{\eta_Z^4 \pi^2 \sqrt{\pi } s\Omega^B }{2 \left|\langle U\rangle_T\right|}  \left ( \frac{\hbar \omega_Z }{k_B T_Z} \right )^{-2} \left( \frac{\langle U\rangle_T}{\sqrt{\pi}  \delta}\right)^7    e^{-  \left(\frac{\langle U\rangle_T}{\sqrt{\pi}  \delta}\right)^2}. \label{tem}
\end{eqnarray}

\begin{figure}[]
{\includegraphics[width=1.\linewidth]{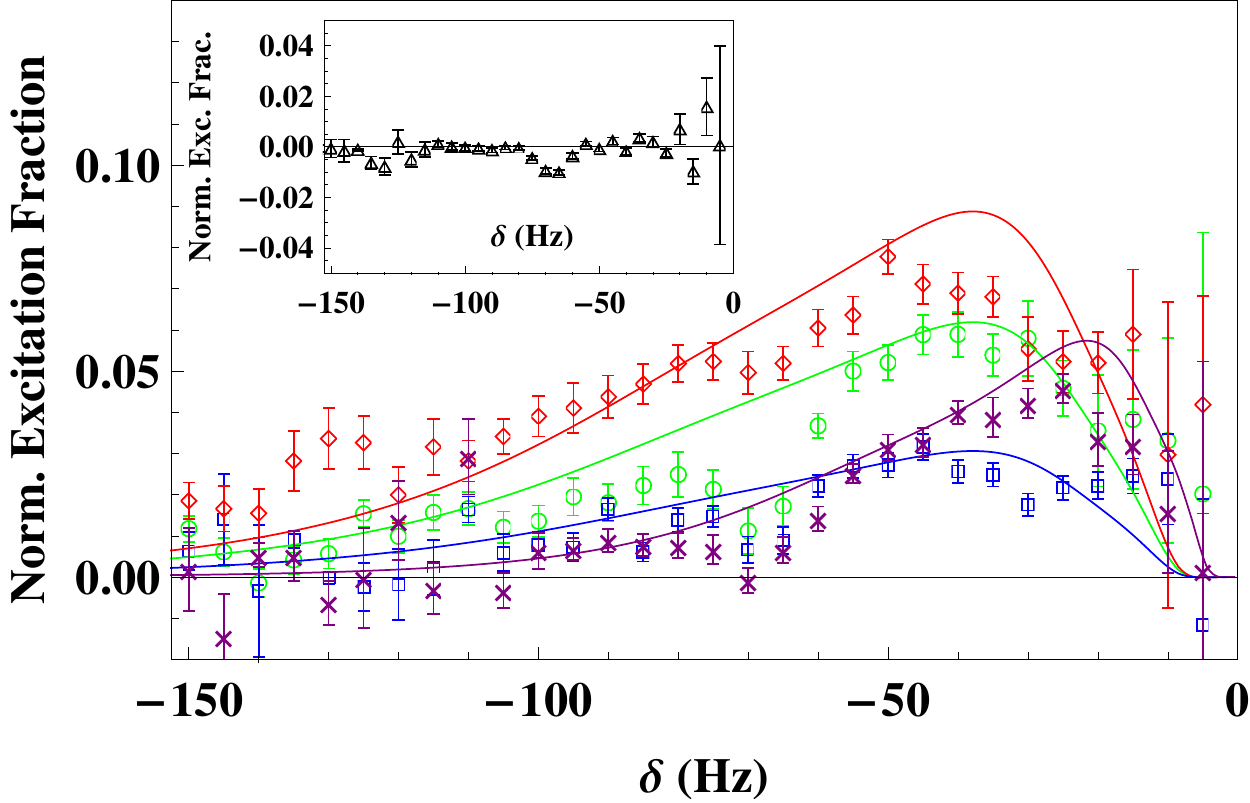}}
 \leavevmode
 \caption{ (Color online) Finite temperature ISB in a 2D lattice ($g$ to $e$): Symbols %(similar conventions to those used in Fig.2)
 correspond to experimental data obtained after reflecting  the binned lineshape at positive  detuning (free of ISB)  and subtracting it from the corresponding bin at negative detuning. Data conditions previously shown in Fig.\ \ref{expe} are displayed with the same symbols and colors. The data indicated by purple crosses corresponds to an 80 ms $\pi$-pulse but with  different trapping conditions ($\omega_{X,Y,Z}= 2 \pi \times \{92,60, 0.6\}$ kHz at trap center,  $T_{X,Y,Z}\sim \{2.5,2.5, 7.2\}$  $\mu$K).  The inset shows the absence of ISB in a 1D lattice. Solid curves correspond to theoretical predictions. { The dip at $\delta\sim -70$ Hz in the measured ISB is due to residual  $m_I=7/2$ atoms.}
 }
\label{lineshape}
\end{figure}

 Fig.\ \ref{lineshape} shows comparisons between  the  measured ISB and theoretical curves calculated using Eq.\ (\ref{tem}). In the 2D lattice, sites near the wings of the lattice beams' Gaussian intensity profiles are significantly shallower than  near the center of the beam intersection region, %, where the trap depths are greatest
 %$\omega_{\perp}(X,Y)\approx \omega_{\perp}(0,0)Exp[- \frac{(X^2+Y^2)}{W_\perp^2}]$
 i.e $\omega_{\perp}(X,Y)\approx \omega_{\perp}(0,0)\textrm{Exp}\!\left[- (X^2+Y^2)/W_\perp^2\right]$, with $\omega_{\perp}\equiv \sqrt{\omega_X \omega_Y}$ and  $W_\perp \equiv \sqrt{W_{X} W_Y}\sim 30$ $\mu$m the mean beam waist.  This implies $u$ varies from tube to tube; so to  compare with experiment, it is important  to average Eq. (\ref{tem}) over the populated tube distribution. We assume a uniform distribution of doubly occupied tubes.
% a   uniform tube population consistent with the loading procedure for the   average   and  scale it  by  the fraction  of the atomic population in doubly occupied lattice sites.
With this procedure,  we obtain lineshapes %which
{that reproduce the  experimental data if we use $a_{eg}^-=-280 a_0$ and $\eta_Z\sim 0.07(0.05)$  for the  $\omega_{\perp}=2\pi \times 87(73)$ kHz cases. The  value of $a_{eg}^-$ that %which
fits the experimental data is outside
 the range of effective scattering lengths $a_{eg}^{-*}\sim -(35-50) a_0$ reported in Ref.\ \cite{Swallows11}. However the discrepancy is attributed to the fact that $a_{eg}^{-*}$ in Ref.~\cite{Swallows11} corresponded to the effective scattering length  required to model the entire lattice array by a single tube at the trap center, and thus $|a_{eg}^{-*}|$  should be smaller than $|a_{eg}^{-}|$.}
   In agreement with theory,  the ISB's position and width are on the order of %determined by
   $\langle U\rangle_{T}$ and their resonant peak heights (normalized to the carrier)   decrease   with decreasing $\Omega^B$ and  increase  with pulse area $s \pi$.
   %{\color{green}Tunneling effects neglected in our model could in principle  modify the shape and %height of the ISB  and  generate additional frequency shifts in the carrier. However, the fact    %we can reproduce (up to an overall scaling factor)  all the data  using  a single scattering %length, after performing  thermal averages and after taking into account  the tube distribution, %regardless of  significantly different  tunneling rates in the various cases, strongly suggest   %our  tunneling-free model is capturing the correct physics. The scaling issue, connected to the %fact that we need different $\eta_Z$  to match the observed peak height can not ne taken very %seriously due to  our normalization procedure and the possibility that different trapping %conditions could give rise indeed to different $\eta_Z$. Nevertheless, we can not rule out   %tunneling effects in the  $T_X=T_y=4.5\mu$K case to be  partly responsible for this scaling  %issue.}

{We determine the nuclear spin purity of the atomic sample to be
$>97\%$ by scanning the probe laser over the clock transition frequencies
for other nuclear spin states. The signature of
the remaining $3\%$ is the small resonant peak at $\delta\sim 70$ Hz
in Fig.\ \ref{expe}. By comparing the amplitude of this small peak with the
ISB,  we conclude that  the observed  sidebands are not
caused by $s$-wave collisions between unpolarized atoms.}

We also performed $e$ to $g$ Rabi interrogation by first transferring all the atoms from $g$ to $e$ with a resonant pulse, then removing  the %reminder
remaining $g$ atoms and finally interrogating the $e$ atoms with an $80$ ms $\pi$-pulse. However, even in the tight geometry we did not see evidence of ISB in this case. The lack  of ISB can be attributed to the absence of double occupancy in a single tube, possibly due to a combination of the following reasons.  (1) A decrease in the number of interrogated atoms  due to the additional steps required to prepare the atoms in  $e$, $(2)$ A more favorable transfer from $g$ to $e$ in tubes with a single atom and (3) Inelastic two-body $e-e$ losses due to $p-$wave collisions similar to the ones recently measured in KRb molecules \cite{Ni} { and in ${}^{88}$Sr (s-wave) \cite{Lisdat}}. We measured the $e$ atom lifetime in a 1D lattice to be   $600$ms. The measured  loss rate, rescaled %been scaled
by the corresponding density in the tight confinement case, translates to decay rates about $30$ times faster, i.e.  lifetimes $\sim 20$ ms.  Given that there is a $10$ ms waiting  time between the preparation and the clock interrogation, $e-e$ %$e-$
losses seem to be the relevant mechanism that leads to  the absence of ISB.

The effect of tunneling between tubes could in principle  modify the shape and height of the ISB (but not its mean position) and  generate an additional frequency shift of the carrier. The latter can be particularly relevant % relevant specially
in the strongly interacting regime where collisional  shifts would otherwise be suppressed. %in the isolated tube array  are suppressed.
In order to investigate the effect of tunneling, we also took one set of data under conditions %(see Fig. 2)
which reproduced those used in Ref.\ \cite{Swallows11}.  The presence of ISB in this lineshape, measured with $T_{X,Y}\sim 2.5$  $\mu$K (see Fig.\ \ref{lineshape}), completely rules out the possibility that these spectroscopic features are caused by tunneling since, under these conditions, we predict that tunneling rates are suppressed by two orders of magnitude compared to $T_{X,Y}\sim 4.5$  $\mu$K.

%Nevertheless, at the current experimental conditions and temperature  we estimate using a %simple double well model that tunneling corrections of the frequency shift should  be lower %than $10\%$ and thus negligible in our experiment.

The studies presented here demonstrate that  tools developed for  clock experiments  can provide  precise understanding of strongly  interacting many-body systems.  %{\it Acknowledgements}
This work was supported by ARO DARPA OLE, NIST, NSF, AFOSR, NRC, and Lee A.\ DuBridge and NDSEG fellowships.
%This work was supported by a grant from the ARO with funding from the DARPA OLE program, NIST, the NSF Physics Frontier Center at JILA, NSF (Grant No. PHY-0803371), (NSF-KITP Grant No. PHY05-51164) AFOSR and by the Lee A. DuBridge Fellowship.  M. Bishof is supported by an NDSEG graduate fellowship, M. Swallows by NRC.

\clearpage

\appendix*

\section{Supplemental material}

\subsection{Interaction parameters}
We consider    nuclear-spin polarized fermionic atoms with two different electronic degrees of freedom ($e,g$) confined in a harmonic trap with trapping frequencies $\omega_{X,Y,Z}$ and assume the atoms interact only via s-wave interactions. The  matrix element that describes  collisions between two atoms ($i = 1, 2$) in the $\textbf{n}_{i} = (n_{iX},n_{iY},n_{iZ})$ harmonic oscillator eigenmodes $\phi_{n_{iX}}^{X}(X) \phi_{n_{iY}}^{Y}(Y) \phi_{n_{iZ}}^{Z}(Z)$ is given by
%The  matrix element that describes  collisions between two atoms in the  $\{\textbf{n}_1,\textbf{n}_2\}$  harmonic oscillator eigenmodes $\phi_{n_{iX}}^{X}(X)$, $\phi_{n_{iY}}^{Y}(Y)$, and $\phi_{n_{iZ}}^{Z}(Z)$ [here $i=1, 2$ and $\textbf{n}_{i} = (n_{iX},n_{iY},n_{iZ})$] is given by
\begin{eqnarray}
%U_{\bf{n}_1,\bf{n}_2} &=&\frac{8 \pi \hbar a_{eg}^-}{m}\int  d^3\mathbf{R} \prod_{i=X,Y,Z} [\phi_{n_{1\mathbf{R}_i}}^{\mathbf{R}_i}(\mathbf{R}_i)]^2 [\phi_{n_{2\mathbf{R}_i}}^{\mathbf{R}_i}(\mathbf{R}_i)]^2 \notag\\
U_{\bf{n}_1,\bf{n}_2} &=&\frac{8 \pi \hbar a_{eg}^-}{m}\int  d^3\mathbf{R} \prod_{\mu =X,Y,Z} [\phi_{n_{1\mu}}^{\mu}(\mu)]^2 [\phi_{n_{2\mu}}^{\mu}(\mu)]^2 \notag\\
& = & u I_{n_{1X},n_{2X}}I_{n_{1Y},n_{2Y}} I_{n_{1Z},n_{2Z}},
\end{eqnarray}
%with
where $\mathbf{R} = (X,Y,Z)$, $u = 4  a_{eg}^- \sqrt{\frac{m \omega_X \omega_Y \omega_Z}{h}}$, %$I_{n_1,n_2}=\frac{\sqrt{2 \pi} \int e^{-2\xi^2 }H_{n_1}^2(\xi) H_{n_2}^2(\xi) d\xi}{2^{n_1+n_2} (n_1!) (n_2!)}$
$I_{n_1,n_2}=\frac{\sqrt{2} \int e^{-2 z^2 }H_{n_1}^2(z) H_{n_2}^2(z) d z}{\sqrt{\pi} 2^{n_1+n_2} (n_1!) (n_2!)}$,
%$I_{n_1,n_2}=\sqrt{2} \int e^{-2 z^2 }H_{n_1}^2(z) H_{n_2}^2(z) d z/\left[\sqrt{\pi} 2^{n_1+n_2} (n_1!) (n_2!)\right]$,
and %$H_{n}(\xi)$
$H_{n}(z)$  are Hermite polynomials.

At finite temperature, expectation values need to be calculated by averaging  over all possible  combinations of modes
weighted  according to their  Boltzmann factor:  $\langle U\rangle_{T} \approx u  \vartheta(\frac{\hbar\omega_X }{ k_B T_X})\vartheta(\frac{\hbar\omega_Y }{ k_B T_Y})\tilde \vartheta(\frac{\hbar\omega_Z }{ k_B T_Z})$ with
%$\vartheta(\alpha)\equiv\left( \frac{\sum_{n_{1},n_{2}}I_{{n_1},{n_2}} e^{-\alpha (n_1+n_2 )}}{\sum_{n_{1},n_{2}} e^{-\alpha(n_1+n_2) }}\right)$
$\vartheta(\alpha)\equiv \frac{\sum_{n_{1},n_{2}}I_{{n_1},{n_2}} e^{-\alpha (n_1+n_2 )}}{\sum_{n_{1},n_{2}} e^{-\alpha(n_1+n_2) }}$
and %$\tilde \vartheta(\alpha)\equiv\left( \frac{\sum_{n_{1}>n_{2}}I_{{n_1},{n_2}} e^{-\alpha (n_1+n_2 )}}{\sum_{n_{1}>n_{2}} e^{-\alpha(n_1+n_2) }}\right)$.
$\tilde \vartheta(\alpha)\equiv \frac{\sum_{n_{1}>n_{2}}I_{{n_1},{n_2}} e^{-\alpha (n_1+n_2 )}}{\sum_{n_{1}>n_{2}} e^{-\alpha(n_1+n_2) }}$. The $\tilde \vartheta$ function is introduced since in our experiment  atoms start polarized in $g$, and  even at $T=0$  they  are  required to occupy  different modes  along $\hat Z$ to satisfy the Pauli exclusion principle.

For $\frac{\hbar\omega_Z }{ k_B T_Z} \ll 1$, eigenmodes with  $|n_{1Z}- n_{2Z}|\gg 1$ dominate, so that
%making Eq.\ (S2) valid for $I_{n_{1Z},n_{2Z}}$. For $|n_{1}- n_{2}| > 1$, to a good approximation, %to a very good approximation, After a close analysis of   $I_{n_1,n_{2}}$ one can show that for $|n_{1}- n_{2}| \gg 1$, to a very good approximation,
\begin{eqnarray}
I_{n_{1Z},n_{2Z}}\approx 2 \frac{K\left[\frac{1}{2} \left(1-\frac{n_{1Z}+n_{2Z}}{|n_{1Z}-n_{2Z}|}\right)\right]}{\pi\sqrt{ \pi |n_{1Z}-n_{2Z}|}},
\end{eqnarray}
where $K$ is the complete elliptic integral of the first kind. %EllipticK  Integral function.
Furthermore, we checked numerically that, for our parameters, ISB is not significantly perturbed by approximating $K$ with $K(0)$, so that %$I_{|n_1- n_2|\gg 1}\approx \frac{1}{\sqrt{\pi }  \sqrt{|n_1-n_2|}}$
 $I_{n_{1Z},n_{2Z}} \approx \frac{1}{\sqrt{\pi }  \sqrt{|n_{1Z}-n_{2Z}|}}$.
 %Since for $\frac{\hbar\omega_Z }{ k_B T_Z}<1$ terms with $|n_1- n_2|\gg 1$ are dominant one can use this approximation to show that
%Using this approximation for $I_{n_{1Z},n_{2Z}}$,
Using this approximation, one can show that
$\vartheta(\alpha \ll 1)\sim\tilde \vartheta(\alpha \ll 1)\sim \sqrt{\alpha}$.

The (absolute value of the) mean interaction energy per particle $E_{\rm int}$ is
$(N-1)|\langle U\rangle_{T}|/2$, where
$N$ is the number of atoms in the trap.

\subsection{Interaction sidebands (ISB) at finite Temperature  for two atoms (N=2)}
Here, we derive Eq.\ (3) in the main text by performing a thermal average over different axial modes.  First, consider a given set of populated modes along $\hat Z$, $\vec{n}=(n_1,n_2)$. If the atoms are in the strongly interacting regime,
$|{U}_{\vec{n}}| \gg \bar{\Omega}_{\vec{n}}$, the ISB  is given by
\begin{eqnarray}
 N^{e}_{\vec{n}}(t,\delta) &\approx&  f(t,\delta -U_{\vec{n}}, \Delta{\Omega}_{\vec{n}}) \label{R}
\end{eqnarray}
with $f(t,\delta, y)\equiv  \frac{y^2}{y^2+\delta^2} \sin^2\left (\frac{t \sqrt{ y^2+\delta^2}}{2}\right)$.

At finite temperature, we need to evaluate $\langle N^e \rangle_T = \frac{\sum_{n_{1}>n_{2}}N^{e}_{\vec{n}}(t,\delta)e^{-\alpha (n_1+n_2 )}}{\sum_{n_{1}>n_{2}} e^{-\alpha(n_1+n_2) }}$ with $\alpha= \frac{\hbar\omega_Z }{ k_B T_Z}$. %\hbar \omega_Z/(k_B T_Z)$.
 %is a complicated sum but
We evaluate $\langle  N^{e} \rangle_T$ under the approximations $
\Delta{\Omega}_{\vec{n}}\approx \frac{\Omega_B \eta_Z^2 (n_1-n_2)}{\sqrt 2}$ and $U_{\vec{n}}\approx \vartheta(\frac{\hbar\omega_X }{ k_B T_X})\vartheta(\frac{\hbar\omega_Y }{ k_B T_Y})\frac{u}{ \sqrt{\pi|n_1-n_2|}}\equiv  \frac{\xi}{\sqrt{\alpha|n_1-n_2|}} $.

To carry out the sum,  we  treat $\alpha (n_1-n_2)\equiv x$ and $\alpha (n_1+n_2) \equiv y$ as continuous variables  (valid if  $\alpha \ll 1$). We also define $\kappa(x)=\frac{\Omega_B \eta_Z^2 x}{\alpha\sqrt 2}\equiv b x$ and $U(x)=\frac{\xi}{\sqrt{x}}$. Then  $\langle N^e \rangle_T\approx \int_0^{\infty} N^e(x/\alpha) e^{-x} dx$ with $N^e(x/\alpha)=\frac{\kappa^2( x)}{\kappa^2( x)^2+\delta^2} \sin^2\left (\frac{t \sqrt{ \kappa^2( x)+(\delta-U(x)) ^2}}{2}\right)$. Noticing that  $ N^{e}$ is sharply peaked at $x_0=(\xi/\delta)^2$, we evaluate $e^{-x}$ and $\kappa$ at $x = x_0$. We write $x = x_0 + x'$ and expand $U$ in $x'$ to first order.  We then define $p = |\xi| x'/(2 b x_0^{5/2})$, so that
\begin{eqnarray}
\langle N^e \rangle_T & \approx &e^{-x_0} \frac{2 b x_0^{5/2}}{|\xi|}\times \notag\\
&&\int_{-\infty}^{\infty} d p \frac{1}{1 + p^2} \sin^2 \left(c \sqrt{1 + p^2}\right),
\end{eqnarray}
where $c = t b x_0/2$. For $c \ll 1$, the integral in Eq.\ (S4) is approximately equal to $\pi c$. This yields  Eq.\ (3) in the main text.

\end{document}